\documentstyle[11pt]{article}
\begin{document}
\title{Reduced phase space quantization of massive vector theory}
\author{H.-P. Pavel\\
        Fachbereich Physik der Universit\"at, D-18051 Rostock, Germany\\
	pavel@darss.mpg.uni-rostock.de\\[1cm]
        V.N. Pervushin\\
        Bogoliubov Laboratory of Theoretical Physics\\
        Joint Institute for Nuclear Research, 141980 Dubna, Russia\\
	pervush@thsun1.jinr.dubna.su}
\date{\today}
\maketitle

\begin{abstract}
We quantize massive vector theory in such a way that it has a well-defined
massless limit. In contrast to the approach by St\"uckelberg where ghost
fields are introduced to maintain manifest Lorentz covariance, we use
reduced phase space quantization with nonlocal dynamical variables which
in the massless limit smoothly turn into the photons, and check explicitly
that the Poincare algebra is fullfilled. In contrast to conventional
covariant quantization  our approach leads to a propagator which has no
singularity in the massless limit and is well behaved for large momenta.
For massive QED, where the vector field is coupled to a conserved fermion
current, the quantum theory of the nonlocal vector fields is shown to be
equivalent to that of the standard local vector fields.
An inequivalent theory, however, is obtained when the reduced nonlocal massive
vector field is coupled to a nonconserved classical current.

\end{abstract}
\newpage

\section{Introduction}

Massless theories, such as gauge theories, are well-known to be plagued
by infrared singularities.
A possible, very clean way to handle the infrared catastrophe of
gauge theories is to give the gauge field a small mass and take the massless
limit after the observable quantities have been calculated.
The quantization of the corresponding massive vector theory in such
a way that it has a well defined massless limit is therefore
important for investigations of the infrared behaviour of
gauge theories, such as the formation of Lorentz and gauge invariant
condensates.
But a well defined massless limit is not straightforward because the
covariant propagator of massive vector theory \cite{Wei,Itz}
\begin{equation}
D_{\mu\nu}(q)=
-{1\over q^2-M^2}\left(g_{\mu\nu}-{q_\mu q_\nu \over M^2}
\right)~.
\label{Photprop}
\end{equation}
has a singularity at M=0. Furthermore the second term leads to ultraviolet
divergences for large momenta. However, when the vector field is coupled to
a conserved current (like in massive QED) $(q_{\mu}J^{\mu}=0)$ the badly
behaved second term drops out and the theory becomes renormalizable.

Already in 1957, St\"uckelberg \cite{Stu} has proposed
a Lagrangian for massive QED that can be quantized in such a way
that it has a well defined massless limit.
For this purpose he introduced additional
ghost fields in order to maintain manifest Lorentz covariance.
The cost of introducing these spurious degrees of freedom is that one
finally has to verify that the observable
quantities are not effected in the massless limit. This can be very
difficult for complicated theories such as massive non-Abelian Yang-Mills
theories \cite{Ba1}.

We shall show in this paper that in the framework of reduced phase space
quantization \cite{HeP}-\cite{Gog} a quantum theory of massive vector fields
with a well defined massless
limit can be obtained, without the introduction of any additional
unphysical degrees of freedom.
In the scheme of reduced phase space quantization one first eliminates
all unphysical degrees of freedom from the classical theory and then
quantizes only the physical variables.

For massless QED reduced phase space quantization
corresponds to quantization in the Coulomb gauge. Here only the two
transverse components of the gauge field and the fermion field are quantized.
Since through this reduction the manifest Poincar\'e invariance
of the theory is lost, it is necessary to check whether the Poincar\'e
algebra is satisfied both on the classical and on the quantum level.
Zumino \cite{BZu} has proven that QED in the Coulomb gauge
satisfies the quantum Poincar\'e algebra
and has shown that the quantum Lorentz transformation properties
of the transverse gauge fields agree with the corresponding classical ones.

Massive vector theory contains three dynamical degrees of freedom.
As for the case of a massless gauge field the Lagrangian of massive vector
theory does not contain the time derivative of the time component $V_0$
of the vector field. The Euler-Lagrange equation for $V_0$ is thus
not a dynamical equation of motion but a constraint.
The reduced Lagrangian obtained by eliminating
$V_0$ from the Lagrangian using the constraint equation
then contains the three spatial components of the vector field and
their time derivatives.
The question which arises then is about the choice of the dynamical degrees of
freedom. For the solution of the above mentioned problem of a well defined
massless
limit we shall use instead of the three local components $V_k$ (k=1,2,3)
the nonlocal fields
\begin{equation}
V^R_k[\vec{V}]\equiv R_{kj}V_j \equiv
 \left(\delta^T_{kj} -{M^2 \over \vec{\partial}^2
-M^2}\delta_{kj}^{||}\right)V_j
\end{equation}
with the longitudinal and transverse projection operators
\begin{equation}
\label{projop}
\delta_{ij}^{||}\equiv {\partial_i\partial_j \over \vec{\partial}^2}~,
 \ \ \ \ \ \ \
\delta_{ij}^T\equiv \delta_{ij}-\delta_{ij}^{||}~.
\end{equation}
We call the fields $V_k^R$ ``reduced'' fields. In the massless limit
the operator $P$ becomes a projection operator onto the transverse
components of the gauge fields.
We shall show in this paper that the reduced massive vector fields
satisfy the quantum Poincar\'e algebra and
have the same Lorentz transformation properties in the massless limit as
the photons. Furthermore we shall obtain the corresponding propagator
for the reduced fields
\begin{equation}
\label{nonlprop}
D^R_{\mu\nu}(q) =
\delta_{\mu 0}\delta_{\nu 0}{1\over (\vec{q}^2+M^2)}
+\delta_{\mu i}\delta_{\nu j}
\left(\delta_{ij}-{q_iq_j\over (\vec{q}^2+M^2)}\right){1\over q^2-M^2}~,
\end{equation}
which is nonsingular in the massless limit and at the same time well
behaved for large momenta.
It includes an instantaneous Yukawa potential and turns into the
Coulomb gauge propagator in the massless limit.
Reduced phase space quantization in terms of reduced
fields is the $M\neq 0$ generalization of Coulomb gauge quantization
of QED, just as the St\"uckelberg approach is the corresponding
generalization of covariant quantization of QED in the Lorentz gauge.
For the case when the vector field is coupled to a conserved
current $(q_{\mu}J^{\mu}=0)$, as in QED,
the two ways of quantization, local and nonlocal, lead to equivalent
quantum theories,
$J^{\mu}D^R_{\mu\nu}J^{\nu}=J^{\mu}D_{\mu\nu}J^{\nu}$.
When the massive vector field however is coupled to a nonconserved
classical current, the two alternative ways of quantization,
local and nonlocal, will lead to inequivalent theories.

Our paper is organized as follows: In Section II we perform the reduction
of the Lagrangian for free massive vector theory and
investigate the Lorentz transformation properties of the local
as well as of the nonlocal reduced quantum fields.
In Section III  we discuss massive QED, where the massive
vector field is coupled to a conserved fermion current, and show
that the quantization in terms of the nonlocal reduced fields is equivalent
to the conventional quantization in terms of local fields.
In Section IV we show that massive vector theory, when coupled to
a nonconserved classical current, may yield inequivalent quantum theories
for the two alternative scenarios of quantization.
In Section V we give our conclusions and outline possible applications.
In Appendix A the classical Lorentz transformation properties are
briefly discussed.


\section{Free massive vector theory}

\subsection{Reduction of the Lagrangian}

The classical action of the free massive vector theory is given by
\begin{equation}
W[V_{\mu}]=\int d^4x{\cal L}(x)
=\int d^4x\left[-{1\over 4}F_{\mu\nu}F^{\mu\nu}+{1\over 2}M^2V_\mu^2\right]
\label{Lem}
\end{equation}
with the Lorentz , but not gauge invariant Lagrangian ${\cal L}(x)$
and the field strength tensor
 $F^{\mu\nu}=\partial^{\mu}V^{\nu}-\partial^{\nu}V^{\mu}$.
The mass term explicitly breaks gauge invariance. In the limit M=0 it becomes
the gauge invariant Lagrangian of free electromagnetism.
In components the Lagrangian reads
\begin{equation}
{\cal L}(x)= {1\over 2}\left(E_i^2[V_0,\vec{V}]-B_i^2[\vec{V}]
+M^2V_0^2-M^2V_i^2\right)~,
\end{equation}
with the ``electric'' and ``magnetic'' fields
\begin{eqnarray}
E_i[V_0,\vec{V}]&\equiv &F_{0i}=-\dot{V}_i-\partial_iV_0~,\nonumber\\
B_i[\vec{V}]&\equiv &-{1\over 2}\epsilon_{ijk}F_{jk}
=\epsilon_{ijk}\partial_j V_k~,
\end{eqnarray}
where the dot denotes the time derivative.

As in the case of massless electrodynamics, one of the Euler-Lagrange
equations, namely that corresponding to $V_0$ is a constraint rather
than an equation of motion. In the method of reduced phase space quantization
well known for gauge theories \cite{HeP}-\cite{Gog}, which we shall in this
paper apply to massive vector theory, one first eliminates the constraint
variable $V_0$ from the Lagrangian before quantization of the theory.
In QED this naturally leads to a reduction of the variables to the
physical transverse photons. In the following we shall investigate
to which physical variables this reduction leads for the case of massive
vector theory.

The Euler-Lagrange equation for $V_0$ is a constraint
\begin{equation}
{\delta W\over \delta V_0}=0 \ \ \leftrightarrow \ \
(\vec{\partial}^2-M^2)V_0=-\partial_i\dot{V}_i~.
\end{equation}
and corresponds in the massless limit to Gauss law.
It can be solved for $V_0$ as
\begin{equation}
\label{V_0V_i}
V_0 [\vec{V}]=-{1\over \vec{\partial}^2-M^2}\partial_i\dot{V}_i
\end{equation}
and inserted into the original Lagrangian.
Here we have abbreviated
\begin{equation}
{1\over \vec{\partial}^2-M^2}f(\vec{x})\equiv -{1\over 4\pi}\int d^3\vec{y}~
{e^{-M|\vec{y}-\vec{x}|}\over |\vec{y}-\vec{x}|} f(\vec{y})~.
\end{equation}
The electric field $E$ then becomes
\begin{equation}
E_k [\vec{V}]= -R_{kj}\dot{V}_j
\end{equation}
with the reduction operator
\begin{equation}
R_{ij}\equiv
\delta_{ij}-{\partial_i\partial_j\over \vec{\partial}^2-M^2}=
\delta_{ij}^T-{M^2\over \vec{\partial}^2-M^2}\delta_{ij}^{||}
\label{Projop}
\end{equation}
where we have used the longitudinal and transverse projection operators
(\ref{projop}).
In contrast to the massless case, $R_{ij}$ is not a projection operator,
 $R^2\neq R$, but
\begin{equation}
\label{R^2}
R_{ij}R_{jl}=R_{il}+
{M^2\partial_i\partial_l\over \left(\vec{\partial}^2-M^2\right)^2}
\end{equation}
and $R_{ij}$ is invertible
\begin{equation}
R_{ij}^{-1}=\delta_{ij}^T-{\vec{\partial}^2-M^2\over M^2}\delta_{ij}^{||}=
\delta_{ij}-{\partial_i\partial_j\over M^2}~.
\end{equation}
In the massless limit however the reduction operator becomes the
transverse projection operator for photons and ceases to be invertible.

The reduced action
\begin{equation}
W_{\rm red}[\vec{V}]\equiv W [V_{\mu}]\bigg|_{V_0=V_0[\vec{V}]}
\equiv\int d^4x {\cal L}_{\rm red}
\end{equation}
therefore reads
\begin{equation}
W_{\rm red}[\vec{V}]=
\int d^4x {1\over 2}\left(\dot{V}_iR_{ij}\dot{V}_j
-V_i\left(\vec{\partial}^2-M^2\right)R_{ij}V_j\right)~,
\end{equation}
where in the second line we have used partial integration and formula
(\ref{R^2}).

Massive vector theory has three dynamical variables.
A possible choice are the three local spatial fields $V_k$
and their canonical conjugate momenta
\begin{equation}
\label{canmoml}
\Pi_k\equiv{\delta{\cal L}\over \delta \dot{V}_k(x)}
={\delta{\cal L}_{\rm red}\over \delta \dot{V}_k(x)}=-E_k=R_{kj}\dot{V}_j~.
\end{equation}
As mentioned above they yield a propagator that leads to a renormalizable
theory only when coupled to conserved currents.

In terms of the transverse and longitudinal parts
$V_i^{T}\equiv \delta_{ij}^{T}V_j$
and $V_i^{||}\equiv \delta_{ij}^{||}V_j$
\begin{equation}
W_{\rm red}[\vec{V}^{T},\vec{V}^{||}]
=\int d^4x ({\cal L}_{\rm red}^T+{\cal L}_{\rm red}^{||})
\end{equation}
with
\begin{eqnarray}
{\cal L}_{\rm red}^T &=& {1\over 2}\left(\dot{V}_i^{T2}+
V_i^T(\vec{\partial}^2-M^2)V_i^T\right)~,\nonumber\\
{\cal L}_{\rm red}^{||} &=& -{1\over 2}M^2\left(\dot{V}_i^{||}{1\over
\vec{\partial}^2-M^2}\dot{V}_i^{||}+V_i^{||2}\right)~.
\end{eqnarray}
One can see that the massless limit exists and leads to the reduced
Lagrangian of free electromagnetism with the two transverse gauge fields
$A_k^{T}$ representing the physical variables. For this case of
massless QED, Zumino \cite{BZu} has proven that the transverse gauge fields
fulfill the quantum Poincare algebra and has determined their Lorentz
transformation properties. The Lorentz transformation properties of
quantum and classical fields coincide only for the nonlocal
$A_k^T[V]=\delta^T_{kj}A_j$, but not for the local $A_k$.

In view of formula (\ref{Projop}) for the electric field
another possible choice is the nonlocal ``physical'' fields
\begin{equation}
V^R_k[\vec{V}]\equiv R_{kj}V_j =
 \left(\delta^T_{kj} -{M^2 \over \vec{\partial}^2
-M^2}\delta_{kj}^{||}\right)V_j~.
\end{equation}
The nonlocal $V_i^R$ differ from the local $V_i$ in that their
longitudinal components are shortened relative to those of $V_i$.
In the massless limit they are shortened to zero or projected out
and reduce to the transverse gauge fields. We shall therefore
call these nonlocal $V_i^R$ reduced fields.
In terms of these nonlocal variables the action is
\begin{equation}
W[\vec{V}^R]=\int d^4x {\cal L}_{\rm red}
=\int d^4x {1\over 2}\left(\dot{V}^R_iR_{ij}^{-1}\dot{V}^R_j
-V^R_i\left(\vec{\partial}^2-M^2\right)R_{ij}^{-1}V^R_j\right)~.
\end{equation}
The conjugate momenta are
\begin{equation}
\Pi_k^R\equiv{\delta{\cal L}_{\rm red}\over \delta \dot{V}_k^R(x)}=
R^{-1}_{kj}\dot{V}_j^R~.
\end{equation}

In the following section we would like to investigate the Lorentz
transformation properties of the massive fields $V_k$ and $V_k^R$.
We shall show that for both choices the equations of motion and the
Lorentz transformation properties of the classical and quantum fields
coincide. Only the nonlocal reduced $V^R_k$, however, have the same
Lorentz transformation properties in the massless limit as the gauge fields.

\subsection{Lorentz Transformation Properties}

We would now like to quantize the classical theory in such a way that we
have an exact correspondence between the classical and quantum fields with
respect to both their equations of motion and their Lorentz transformation
properties.
In order to investigate the Lorentz transformation properties of the fields
it is necessary to construct the energy momentum tensor
corresponding to the Lagrangian (\ref{Lem}).
Using Noether's theorem one finds the canonical energy momentum tensor
\begin{equation}
T^{\mu\nu}=-F^{\mu\sigma}\partial^\nu V_\sigma - g^{\mu\nu}{\cal L}~,
\end{equation}
which is conserved, $\partial_\mu T^{\mu\nu}=0$, but asymmetric.
The tensor $T_{\mu\nu}$ is determined only up to the total derivative
$\partial_\lambda t^{\lambda\mu\nu}$
with $t^{\lambda\mu\nu}=-t^{\mu\lambda\nu}$ antisymmmetric in the indices
$\lambda$ and $\mu$.
Adding to $T^{\mu\nu}$ the total derivative
\begin{equation}
\partial_\sigma (V^\nu F^{\mu\sigma})=(\partial_\sigma V^\nu)F^{\mu\sigma}
+V^\nu\partial_\sigma F^{\mu\sigma}=(\partial_\sigma V^\nu)F^{\mu\sigma}
+M^2 V^\nu V^\mu~,
\end{equation}
where in the second equality the classical equations of motion
$\partial_\mu F^{\mu\sigma}+M^2 V^\sigma=0$
have been used, one obtains the Belinfante tensor
\begin{equation}
\label{Belimn}
\Theta^{\mu\nu}={F^{\mu}}_{\sigma}F^{\sigma\nu} +{1\over 4}
g^{\mu\nu}F_{\sigma\rho}F^{\sigma\rho}
+M^2(V^\mu V^\nu - {1\over 2}g^{\mu\nu}V_\sigma^2).
\end{equation}
It is symmetric and becomes gauge invariant in the massless limit
$M\rightarrow 0$.
The corresponding angular momentum tensor,
${\cal M}^{\sigma\mu\nu}=x^{\mu}\Theta^{\sigma\nu}-x^{\nu}\Theta^{\sigma\mu}$,
is another conserved tensor current,
$\partial_\sigma {\cal M}^{\sigma\mu\nu}=0$.
From $\Theta_{\mu\nu}$ and ${\cal M}^{\sigma\mu\nu}$
we find the following conserved Poincare charges:
The Hamiltonian
\begin{equation}
\label{PoH}
H=\int d^3\vec{x}~ \Theta_{00}(\vec{x},t)~,
\end{equation}
the 3-momentum
\begin{equation}
\label{PoP}
R_k=\int d^3\vec{x}~ \Theta_{0k}(\vec{x},t)~,
\end{equation}
the angular momenta
\begin{equation}
\label{PoJ}
J_k={1\over 2}\epsilon_{kij}\int d^3\vec{x}{\cal M}_{0ij} =
\epsilon_{kij}x_iR_j+\epsilon_{kij}\int d^3\vec{y}(y_i-x_i)\Theta_{0j}
(\vec{y},t)~,
\end{equation}
and the Lorentz boosts
\begin{equation}
\label{PoK}
K_k=\int d^3\vec{x}{\cal M}_{00k}^B =x_kH-tR_k+
\int d^3\vec{y}(y_k-x_k)\Theta_{00}(\vec{y},t)~.
\end{equation}
All can therefore be obtained from the four components $\Theta_{00}$ and
$\Theta_{0k}$.

Replacing in $\Theta_{\mu\nu}$ the $V_0$ in terms of the spatial $V_k$
via (\ref{V_0V_i}) according to our reduction method
we obtain the reduced components
\begin{equation}
\Theta_{\mu\nu}^{\rm red}\equiv\Theta_{\mu\nu}\Big|_{V_0=V_0[\vec{V}]}~.
\end{equation}
In the following we shall investigate both the quantization in the local
and in the nonlocal reduced fields. We have to check whether
the Poincar\'e algebra is fulfilled in both cases. For convenience
we shall in the remainder of the paper drop the index
``${\rm red}$'' from the $\Theta_{\mu\nu}$.
The corresponding classical Lorentz transformation properties of both the
local and the nonlocal fields are discussed in Appendix A.

\subsection{Quantization in the local $V_k$}

In this section we shall briefly review the standard quantization in the
local fields $V_k$ and $\Pi_k$ and the fulfillment of the corresponding
Poincar\'e algebra.
For a more detailed discussion see e.g. \cite{Wei}.

The local fields, $V_k$, and their canonical conjugate momenta, $\Pi_k$,
are quantized by imposing the canonical commutation relations
\begin{equation}
i[\Pi_k(\vec{x},t),V_j(\vec{y},t)]=\delta_{kj}\delta(\vec{x}-\vec{y})
\end{equation}
with all other commutators vanishing.
In terms of $V_i$ and $\Pi_i$ the four relevant components of the Belinfante
tensor are obtained from (\ref{Belimn}) by replacing $V_0$ in terms of the
spatial $V_k$ via (\ref{V_0V_i}) and
then replacing the velocities $R_{ij}\dot{V}_j$
by the momenta $\Pi_i$ according to (\ref{canmoml})
\begin{eqnarray}
\label{Beli00}
\Theta_{00} &=&{1\over 2}\left(\Pi_i^2+{1\over M^2}(\partial_i\Pi_i)^2
+B_i[\vec{V}]^2+M^2V_i^2\right) ~,\\
\label{Beli0k}
\Theta_{0k} &=&-\Pi_l\partial_k V_l+\partial_l(\Pi_l V_k)~.
\end{eqnarray}
with the operators taken to be symmetrically ordered
$\Pi_k{\cal A} V_i\equiv {1\over 2}\{\Pi_k,{\cal A} V_i\}$. This is important
for the Poincar\'{e} algebra to be fulfilled as discussed below.
Using (\ref{R^2}) and performing an integration by parts the Hamiltonian is
\begin{equation}
H=\int d^3\vec{x}~{1\over 2}[\Pi_iR_{ij}^{-1}\Pi_j -V_i(\vec{\partial}^2-M^2)
R_{ij}V_j]~,
\end{equation}
and the three momenta
\begin{equation}
P_k=-\int d^3\vec{x}~\Pi_i\partial_k V_i~.
\end{equation}
The generators $J_k$ and $K_k$ of rotations and Lorentz boosts are
\begin{equation}
J_k=\epsilon_{kij}\left(x_iP_j + \Pi_i V_j\right)
-\epsilon_{kij}\int d^3\vec{y}(y_i-x_i)~\Pi_l\partial_j V_l~,
\end{equation}
and
\begin{eqnarray}
K_k &=& x_kH-tP_k\nonumber\\
& &~~+\int d^3\vec{y}(y_k-x_k)
{1\over 2}\left(\Pi_i^2+{1\over M^2}(\partial_i\Pi_i)^2
+B_i[\vec{V}]^2+M^2V_i^2\right)~.
\end{eqnarray}
Straightforward but lengthy calculations show that these satisfy the
Poincar\'{e} algebra:
\begin{eqnarray}
\label{Poincare}
~[J_i,J_{j}] &=& i\epsilon_{ijk}J_k~,\nonumber\\
~[J_i,K_{j}] &=& i\epsilon_{ijk}K_k~,\nonumber\\
~[K_i,K_{j}] &=& -i\epsilon_{ijk}J_k~,\nonumber\\
~[J_i,P_{j}] &=& i\epsilon_{ijk}P_k~,\nonumber\\
~[K_i,P_{j}] &=& iH\delta_{ij}~,\nonumber\\
~[J_i,H]   &=& [P_i,H] = [H,H] = 0~,\nonumber\\
~[K_i,H]   &=& iP_i~.
\end{eqnarray}
The Heisenberg equations read
\begin{eqnarray}
\dot{V}_k(\vec{x},t) &=& i[H,V_k(\vec{x},t)]=R_{kj}^{-1}\Pi_j~,\nonumber\\
\dot{\Pi}_k(\vec{x},t) &=& i[H,\Pi_k(\vec{x},t)]=
(\vec{\partial}^2-M^2)R_{kl}V_l~.
\end{eqnarray}
Inserting the first into the second gives
\begin{equation}
\ddot{V}_k=(\vec{\partial}^2-M^2)V_k~,
\end{equation}
in agreement with the classical Klein-Gordon equation.
Furthermore we have the standard transformation property
$i[R_k,V_i(\vec{x},t)]=-\partial_kV_i$ under spatial translations.
For the behaviour of the operator $V_k$ under infinitesimal
Lorentz boosts we find
\begin{equation}
\delta_L V_k=i\epsilon_i[K_i,V_k]=
\epsilon_i (x_i\partial_t+t\partial_i)V_k + \epsilon_k
{1\over \vec{\partial}^2-M^2}\partial_i\dot{V_i}~,
\label{qLTl}
\end{equation}
in agreement with the classical form discussed in Appendix A.
Straightforward but lengthy calculations show that these satisfy the
Poincar\'{e} algebra (\ref{Poincare}) and in particular give the correct
boost-boost commutator, since we correctly obtain the Schwinger anomalous
commutator
\begin{equation}
[\Theta_{00}(\vec{x},t),\Theta_{00}(\vec{y},t)]=-i\left(
\Theta_{0k}(\vec{x},t)+\Theta_{0k}(\vec{y},t)\right)\partial_k^x
\delta(\vec{x}-\vec{y})~.
\end{equation}
It is important to note that these differ in the massless limit from
those of the photon field which does not transform like a vector.
In the next paragraph we shall repeat the same steps for the nonlocal
reduced fields.

\subsection{Quantization in the nonlocal $V_k^R$}

The nonlocal fields $V_k^R[\vec{V}]$ and their
canonical conjugate momenta, $\Pi_k^R$,
are quantized by imposing the canonical commutation relations
\begin{equation}
i[\Pi_k^R(\vec{x},t),V^R_j(\vec{y},t)]=
\delta_{kj}\delta(\vec{x}-\vec{y})
\end{equation}
with all the other commutators vanishing.
In order to obtain the components of the corresponding Belinfante tensor
$\Theta^R_{\mu\nu}$ in terms of $V_k^R$ and $\Pi_k^R$ we replace everywhere in
(\ref{Beli00}) and (\ref{Beli0k}) $\Pi_k$ by $R_{kj}\Pi_j^R$ and $V_k$ by
$R^{-1}_{kj}V_j^R$
\begin{equation}
\Theta^R_{\mu\nu}[\Pi_k^R, V_k^R]\equiv
\Theta_{\mu\nu}[R_{kj}\Pi_j^R, R_{kj}V_j^R]~,
\end{equation}
or explicitly
\begin{eqnarray}
\Theta_{00}^R(\vec{x},t)&=&
{1\over 2}\Big[(R_{ij}\Pi_j^R)^2+
M^2\left({1\over \vec{\partial}^2-M^2}\partial_i
\Pi^R_i\right)^2
+B_i^2[R_{ij}^{-1}V_j^R]\nonumber\\
& &\ \ \ +M^2\left(R_{ij}^{-1}V_j^R\right)^2\Big]~,\\
\Theta_{0k}^R(\vec{x},t)&=&
-(R_{ij}\Pi^R_j)\partial_k(R^{-1}_{il}V^R_l)+
\partial_i\left((R_{ij}\Pi_j^R)(R_{kj}^{-1}V^R_j)\right)
\end{eqnarray}
Again the operators are taken to be symmetrically ordered.
The Hamiltonian and the momentum operators then read
\begin{eqnarray}
H^R &=& \int d^3\vec{x}~{1\over 2}[\Pi_i^RR_{ij}\Pi_j^R -
V_i^R(\vec{\partial}^2-M^2)R_{ij}^{-1}V_j^R]~,\\
P_k^R&=&-\int d^3\vec{x}~\Pi_i^R\partial_k V_i^R~,
\end{eqnarray}
The rotations $J_k^R$ and Lorentz boosts $K_k^R$ are given in terms of
$\Theta_{0k}^R$ and $\Theta_{00}^R$ by (\ref{PoJ}) and
(\ref{PoK}), respectively.

We obtain the Heisenberg equations
\begin{eqnarray}
\dot{V}_k^R(\vec{x},t) &=& i[H^R,V_k^R(\vec{x},t)]=
R_{kj}\Pi_j^R~,\nonumber\\
\dot{\Pi}_k^R(\vec{x},t) &=& i[H^R,\Pi_k^R(\vec{x},t)]=
(\vec{\partial}^2-M^2)R_{ij}^{-1}V_j^R~,
\end{eqnarray}
and thus again the Klein-Gordon equations:
\begin{equation}
\ddot{V}_k^R=(\vec{\partial}^2-M^2)V_k^R~,
\end{equation}
in formal agreement with the classical case.
Furthermore we find the standard transformation
$i[P_k^R,V^R_i(\vec{x},t)]=-\partial_kV^R_i$ under spatial translations.
For the Lorentz transformation properties we find
\begin{equation}
\label{KVP}
\delta_L V_k^R=i\epsilon_i[K_i^R,V_k^R]=
\epsilon_i (x_i\partial_t+t\partial_i)V_k^R - \partial_k
{1\over \vec{\partial}^2-M^2}\epsilon_i\dot{V}^R_i~,
\label{qLTnl}
\end{equation}
in agreement with the classical form discussed in Appendix A.
In contrast to the local fields,
the Lorentz transformation properties of the nonlocal $V_k^R[V]=R_{kj}V_j$
reduce to those of the photon field in the massless limit.
Comparing these with the corresponding transformation of the local
fields, (\ref{qLTl}), we see that we have consistently
$\delta_LV_k^R~=~R_{kl}(\delta_LV_k)$ in accordance with  $V_k^R=R_{kj}V_j$.
Careful, lengthy calculations show that the generators in the nonlocal
variables also satisfy the Poincar\'{e} algebra, in particular the
boost-boost commutator is correct, since here too
\begin{equation}
[\Theta_{00}^R(\vec{x},t),\Theta_{00}^R(\vec{y},t)]=-i\left(
\Theta_{0k}^R(\vec{x},t)+\Theta_{0k}^R(\vec{y},t)\right)\partial_k^x
\delta(\vec{x}-\vec{y})~.
\end{equation}

In summary, we have confirmed the Poincar\'e invariance of both
ways of quantization, in terms of the local and the nonlocal
reduced fields. All quantum transformation properties are in exact
formal agreement with the corresponding classical ones.
Quantization in terms of the nonlocal reduced fields is therefore
a perfectly acceptible alternative to the standard quantization in terms
of the local fields.
The quantum theory of the nonlocal reduced fields has the very important
property that in the massless limit it smoothly reduces to the photon theory.

\section{Massive QED}

In this section we shall consider the case of massive QED, where the massive
vector fields are coupled to a conserved fermion current. We shall
show that the quantization in terms of the nonlocal fields leads to a
theory which is equivalent to that obtained via conventional quantization
in terms of local vector fields, but leads to a propagator for the nonlocal
vector fields which is well behaved in the massless limit as well as in the
ultraviolet limit of large momenta.

\subsection{Reduction of the Lagrangian}

The classical action of massive QED is
\begin{equation}
\label{LQEDloc}
W=\int d^4x {\cal L}(x)=
\int d^4x \left[-{1\over 4}F_{\mu\nu}F^{\mu\nu}+{1\over 2}M^2V_\mu^2+
\bar\Psi(i\not{\!\partial}-m)\Psi -V_{\mu}J^{\mu}\right]~,
\end{equation}
with the Lorentz, but not gauge invariant Lagrangian ${\cal L}(x)$
and the currents $J_\mu\equiv e\bar\Psi\gamma_\mu\Psi$.
As in the free theory the Euler-Lagrange equation for $V_0$ is a constraint
and not an equation of motion. Within the scheme of reduced
phase space quantization, the constraint is used to eliminate $V_0$
from the Lagrangian.

The Euler-Lagrange equation for $V_0$ reads
\begin{equation}
{\delta W\over \delta V_0}=0 \ \ \leftrightarrow \ \
(\vec{\partial}^2-M^2)V_0=-\partial_i\dot{V}_i+J_0~,
\end{equation}
which in massless QED corresponds to Gauss' law.
It has the solution
\begin{equation}
\label{V_0V_iQED}
V_0[\vec{V},J_0]
={1\over \vec{\partial}^2-M^2}(-\partial_i\dot{V}_i +J_0)~.
\end{equation}
Inserting this into W we obtain the reduced $W_{\rm red}$
\begin{equation}
W_{\rm red}[\vec{V},\Psi]=\int d^4x {\cal L}_{\rm red}
\equiv \int d^4x \left({\cal L}_{\rm red}^{V} +{\cal L}_{\rm red}^\Psi\right)
\label{WredmQED}
\end{equation}
with the reduced Lagrangian
\begin{eqnarray}
{\cal L}^V_{\rm red} &=& {1\over 2}\left(\dot{V}_iR_{ij}\dot{V}_j+
                 V_i(\vec{\partial}^2-M^2)R_{ij}V_j\right)~,\nonumber\\
{\cal L}^\Psi_{\rm red} &=& {1\over 2}J_0{1\over \vec{\partial}^2-M^2}J_0
      +J_0\left({1\over \vec{\partial}^2-M^2}\partial_i\dot{V}_i\right)
                  +V_iJ_i+\bar\Psi(i\not{\!\partial}-m)\Psi~,
\label{Lredpsi}
\end{eqnarray}
with  $R_{ij}$ given by (\ref{Projop}).

As in the case of free massive vector theory discussed in the preceding
section we have several choices for the dynamical variables.
One possibility is to choose the local fields $V_k$ and $\Psi$ and their
canonical conjugate momenta
\begin{eqnarray}
\Pi_i &\equiv &
 {\delta {\cal L}_{\rm red}\over \delta \dot{V}_i}= R_{ij}\dot{V}_j
+{1\over \vec{\partial}^2-M^2}\partial_i J_0~,\nonumber\\
\Pi &\equiv &
 {\delta {\cal L}_{\rm red}\over \delta \dot{\Psi}}= i\Psi^+~.
\label{canmomQED}
\end{eqnarray}

Alternatively we can eliminate the second term in ${\cal L}_{\rm red}^\Psi$
by introducing the new variable
\begin{equation}
\label{psired}
\Psi^R\equiv \exp\left(ie{1\over \vec{\partial}^2-M^2}
                         \partial_iV_i\right)\Psi
\end{equation}
which generalizes the gauge invariant Dirac variables in QED.
Since (\ref{psired}) is only a phase transformation, the corresponding
current $J_{\mu}$ stays the same and
${\cal L}^\Psi_{\rm red}$ becomes
\begin{equation}
{\cal L}^\Psi_{\rm red} = {1\over 2}J_0{1\over \vec{\partial}^2-M^2}J_0
      +J_i\left(\delta_{ij}-{\partial_i\partial_j\over \vec{\partial}^2-M^2}
      \right)V_j +\bar\Psi^R(i\not{\partial}-m)\Psi^R~.
\end{equation}
Using the nonlocal variables $V^R_k=R_{kj}V_j$ with $R_{ij}$ given by
(\ref{Projop}) as for free massive vector theory
the reduced Lagrangian can be written
\begin{eqnarray}
{\cal L}_{\rm red} &=& {1\over 2}\left(\dot{V}^R_iR^{-1}_{ij}\dot{V}^R_j+
               V^R_i(\vec{\partial}^2-M^2)R^{-1}_{ij}V^R_j\right)\nonumber\\
               & & +J_iV^R_i
                   +{1\over 2}J_0{1\over \vec{\partial}^2-M^2}J_0
                  +\bar\Psi^R(i\not{\partial}-m)\Psi^R~.
\label{redL}
\end{eqnarray}
The corresponding canonical conjugate momenta are
\begin{eqnarray}
\Pi^R_i &\equiv & {\delta {\cal L}\over \delta \dot{V}^R_i}
= R^{-1}_{ij}\dot{V}_j^R
\nonumber\\
\Pi^R &\equiv & {\delta {\cal L}\over \delta \dot{\Psi}^R} = i\Psi^{P+}~.
\end{eqnarray}
In the following paragraphs we shall investigate both ways of quantization,
the standard one in terms of the local fields $V_k$ and $\Psi$, and the
alternative one in terms of the nonlocal reduced fields $V^R_k$ and
$\Psi^R$. We shall check the Poincar\'e invariance of both.

\subsection{Lorentz transformation properties}

As in the free massive vector theory the generators of the Poincare algebra
are obtained from the Belinfante tensor.
For massive QED it has the form
\begin{eqnarray}
\label{BelimnQED}
\Theta^{\mu\nu}&=&
{F^{\mu}}_{\sigma}F^{\nu\sigma} +{1\over 4}
g^{\mu\nu}F_{\sigma\rho}F^{\sigma\rho}
+M^2(V^\mu V^\nu - {1\over 2}g^{\mu\nu}V_\sigma^2)\nonumber\\
& &+\overline{\Psi}i\gamma^{\mu}\partial^{\nu}\Psi-J^{\mu}V^{\nu}
 +{i\over 4}\partial_{\lambda}(\overline{\Psi}
\Gamma^{\lambda\mu\nu}\Psi)-g^{\mu\nu}\overline{\Psi}\gamma_{\sigma}
(i\partial^{\sigma}-m)\Psi+ g^{\mu\nu}J_{\sigma}V^{\sigma}~,
\end{eqnarray}
with
\begin{equation}
\Gamma^{\lambda\mu\nu}\equiv
{1\over 2}[\gamma^{\lambda},\gamma_{\mu}]\gamma^{\nu}
+g^{\nu\mu}\gamma^{\lambda}-g^{\lambda\nu}\gamma^{\mu}~.
\end{equation}
From these we obtain the Poincare generators $H,P_k,J_k$, and $K_k$
via the formulae (\ref{PoH})-(\ref{PoK}).

\subsection{Quantization in the local $V_i$ and $\Psi_{\alpha}$}

In this section we shall first review the standard quantization
in terms of the local fields $V_k$ and $\Psi_{\alpha}$,
as discussed e.g. in \cite{Wei}.
The local fields $V_k,\Psi_{\alpha}$ and their canonical conjugate momenta
$\Pi_k,\Pi_{\alpha}$
are quantized by imposing the canonical commutation and anticommutation
relations
\begin{eqnarray}
i[\Pi_k(\vec{x},t),V_j(\vec{y},t)]&=&\delta_{kj}\delta(\vec{x}-\vec{y})~,
\nonumber\\
\{\Pi_{\alpha}(\vec{x},t),\Psi_{\beta}(\vec{y},t)\}
&=&\delta_{\alpha\beta}\delta(\vec{x}-\vec{y})~,
\nonumber\\
\{\Psi_{\alpha}(\vec{x},t),\Psi_{\beta}(\vec{y},t)\}
&=&\{\Pi_{\alpha}(\vec{x},t),\Pi_{\beta}(\vec{y},t)\}=0~,
\end{eqnarray}
with all other commutators vanishing.
The relevant components of the Belinfante tensor in terms of
$V_k,\Psi_{\alpha}$ and $\Pi_k,\Pi_{\alpha}$
are obtained from (\ref{BelimnQED}) by replacing
$V_0$ in terms of the spatial components of the vector field via
(\ref{V_0V_iQED})
and then replacing the velocities $\dot{V}^R,\dot{\Psi}$ in terms of the
momenta $\Pi_k,\Pi$ according to (\ref{canmomQED})
\begin{eqnarray}
\label{theta00lf1}
\Theta_{00} &=&
{1\over 2}\left(\Pi_i^2+{1\over M^2}(\partial_i\Pi_i)^2
+B_i[\vec{V}]^2+M^2V_i^2\right)\nonumber\\
 & &~-\Pi\gamma_0(\gamma_i\partial_i+im)\Psi - J_iV_i
+{1\over 2M^2}J_0^2-{1\over M^2}(\partial_i\Pi_i)J_0
-{1\over 4}\partial_i\left(\Pi[\gamma_i,\gamma_0]\Psi\right)~,\\
\label{theta00lf2}
\Theta_{0k} &=&
-\Pi_i\partial_k V_i
+\partial_i(\Pi_i V_k)\nonumber\\
     & & -\Pi\partial_k\Psi
             +{1\over 8}\partial_i\left(\Pi\left([\gamma_i,\gamma_k]-
4\gamma_i\gamma_k\right)\Psi\right)~,
\end{eqnarray}
with $J_{\mu}=-ie\Pi\gamma_0\gamma_{\mu}\Psi$ in terms of $\Psi$ and $\Pi$.
The bosonic operators are taken to be symmetrically ordered and the fermionic
ones to be ordered antisymmetrically $\bar{\Psi}{\cal M}\Psi\equiv
{1\over 2} [\bar{\Psi},{\cal M}\Psi]$.
The Hamiltonian is
\begin{eqnarray}
H &=& \int d^3\vec{x}\Big\{{1\over 2}\left(\Pi_iR^{-1}_{ij}\Pi_j-
                 V_i(\vec{\partial}^2-M^2)R_{ij}V_j\right)-
\Pi\gamma_0(\gamma_i\partial_i+im)\Psi\nonumber\\ \ \ \ \ \ \
& &\ \ \ \ \ \ \ \
-V_iJ_i + {1\over 2M^2}J_0^2 - J_0{1\over M^2}\partial_i\Pi_i\Big\}~,
\end{eqnarray}
and the momenta are
\begin{equation}
P_k=-\int d^3\vec{x}
\left(\Pi_i\partial_k V_i+\Pi\partial_k\Psi\right)~.
\end{equation}
Hence we obtain the Heisenberg equations
\begin{eqnarray}
\dot{V}_k(\vec{x},t) &=& i[H,V_k(\vec{x},t)]=R_{kj}^{-1}\Pi_j+{1\over M^2}
\partial_kJ_0~,\nonumber\\
\dot{\Pi}_k(\vec{x},t) &=& i[H,\Pi_k(\vec{x},t)]=
(\vec{\partial}^2-M^2)R_{kl}V_l+J_k~.
\end{eqnarray}
Inserting the first into the second gives
\begin{equation}
\ddot{V}_k=(\vec{\partial}^2-M^2)V_k+R_{kj}^{-1}J_j+{1\over M^2}
\partial_k\dot{J}_0~.
\end{equation}
Using the operator current conservation,
$\dot J_0=-\partial_i J_i$, following from the equations of motion
of the fermions, discussed in the following,
we obtain the Klein-Gordon equation with a fermionic source term:
\begin{equation}
\label{locKGeq}
(\Box + M^2)V_k=J_k~.
\end{equation}
For the fermions we have
\begin{equation}
\dot\Psi= i[H,\Psi]=\gamma_0(-i\gamma_i\partial_i-e\gamma_i V_i +m)\Psi
         -{e\over M^2}\left(-\partial_i\Pi_i+J_0\right)\Psi~.
\end{equation}
Defining the operator
\begin{equation}
V_0\equiv {1\over M^2}\left(-\partial_i\Pi_i+J_0\right)
\end{equation}
we find the covariant Dirac equation
\begin{equation}
(i\gamma^{\mu}\partial_{\mu}-e\gamma^{\mu}V_{\mu}-m)\Psi=0~.
\end{equation}
For the behaviour of the operator $V_k$ and $\Psi_{\alpha}$ under
infinitesimal Lorentz boosts we find
\begin{eqnarray}
\delta_L V_k &=& i\epsilon_i[K_i,V_k]=
\epsilon_i (x_i\partial_t+t\partial_i)V_k + \epsilon_k
{1\over \vec{\partial}^2-M^2}
\left(-\partial_i\dot{V}_i+J_0\right)~,\\
\delta_L\Psi &=& i\epsilon_i[K_i,\Psi]=
\epsilon_i(x_i\partial_t+t\partial_i)\Psi+{1\over 4}
\epsilon_k[\gamma_0,\gamma_k]\Psi~.
\end{eqnarray}
Lengthy, but straightforward calculations show that the Poincar\'e algebra
is indeed fulfilled, in particular the boost-boost commutator is obtained
correctly. In the following paragraph we shall repeat all the steps
for the nonlocal reduced fields.

\subsection{Quantization in the nonlocal $V^R_i$ and $\Psi_{\alpha}^R$}

We shall now repeat the same steps for the nonlocal reduced fields
$V_k^R$ and $\Psi^R$.
The nonlocal fields $V^R_k,\Psi^R$ and their canonical conjugate momenta
$\Pi^R_k,\Pi^R$
are quantized by imposing the canonical commutation and anticommutation
relations
\begin{eqnarray}
i[\Pi^R_k(\vec{x},t),V^R_j(\vec{y},t)]&=&\delta_{kj}\delta(\vec{x}-\vec{y})~,
\nonumber\\
\{\Pi^R_{\alpha}(\vec{x},t),\Psi^R_{\beta}(\vec{y},t)\}
&=&\delta_{\alpha\beta}\delta(\vec{x}-\vec{y})~,\nonumber\\
\{\Psi^R_{\alpha}(\vec{x},t),\Psi^R_{\beta}(\vec{y},t)\}&=&
\{\Pi^R_{\alpha}(\vec{x},t),\Pi^R_{\beta}(\vec{y},t)\}=0
\end{eqnarray}
with all other commutators vanishing.
The components of the Belinfante tensor in terms of the nonlocal fields are
obtained from (\ref{theta00lf1}) and (\ref{theta00lf2}) by replacing
$V_k\rightarrow R^{-1}_{kj}V^R_j$ and $\Pi_k\rightarrow R_{kj}\Pi^R_j
+(\partial_k/\vec{\partial}^2-M^2)J_0$
\begin{equation}
\Theta^R_{\mu\nu}[\Pi_k^R, V_k^R]\equiv
\Theta_{\mu\nu}[R_{kj}\Pi_j^R,
                R_{kj}V_j^R+ \partial_k/(\vec{\partial}^2-M^2)J_0]~,
\end{equation}
or explicitly
\begin{eqnarray}
\Theta_{00}^R &=&
{1\over 2}\Big[(R_{ij}\Pi_j^R)^2+
M^2\left({1\over \vec{\partial}^2-M^2}\partial_i\Pi^R_i\right)^2
\nonumber\\
& &\ \ \ +B_i^2[R_{ij}^{-1}V_j^R] +M^2\left(R_{ij}^{-1}V_j^R\right)^2\Big]
-\Pi^R\gamma_0(\gamma_i\partial_i+im)\Psi^R\nonumber\\
& &
-J_iV^R_i
+{1\over 2}\left({1\over \vec{\partial}^2-M^2}\partial_iJ_0\right)^2
+{1\over 2}M^2\left({1\over \vec{\partial}^2-M^2}J_0\right)^2
\nonumber\\
& & -{1\over 4}\partial_i\left(\Pi^R[\gamma_0,\gamma_i]\Psi^R\right)
+\partial_i\left(R_{ij}\Pi_j^R{1\over \vec{\partial}^2-M^2}J_0\right)~,\\
\Theta_{0k}^R &=&
-(R_{ij}\Pi^R_j)\partial_k(R^{-1}_{ij}V^R_j)+
\partial_i\left((R_{ij}\Pi_j^R)(R_{kj}^{-1}V^R_j)\right)
\nonumber\\
& & -\Pi^R\partial_k\Psi^R
 +{1\over 8}\partial_i\left(\Pi^R\left([\gamma_i,\gamma_k]
-4\gamma_i\gamma_k\right)\Psi^R\right) - \partial_i\left(\left(
{1\over \vec{\partial}^2-M^2}J_0\right)\partial_k(R_{il}^{-1}V_l^R)\right)
\nonumber\\
& & +\left(\left({1\over \vec{\partial}^2-M^2}J_0\right) -
J_0{1\over \vec{\partial}^2-M^2}\right)
\partial_i\partial_k (R_{il}^{-1}V_l^R)
\end{eqnarray}
Again the bosonic operators are taken to be symmetrically ordered
and the fermionic ones to be ordered antisymmetrically.
The Hamiltonian is
\begin{eqnarray}
H^R &=& \int d^3\vec{x}\Big\{{1\over 2}\left(\Pi^R_iR_{ij}\Pi^R_j-
                 V^R_i(\vec{\partial}^2-M^2)R^{-1}_{ij}V^R_j\right)
-\Pi^R\gamma_0(\gamma_i\partial_i+im)\Psi^R\nonumber\\
 & &\ \ \ \ \ \ \ \ \ \
-V^R_iJ_i - {1\over 2}J_0{1\over \vec{\partial}^2-M^2}J_0\Big\}
\end{eqnarray}
and the momenta are
\begin{equation}
P_k^R=-\int d^3\vec{x}\left(
\Pi^R_j\partial_k V^R_j +\Pi^R\partial_k\Psi^R\right)~.
\end{equation}
This leads to the Heisenberg equations
\begin{eqnarray}
\dot{V}_k^R(\vec{x},t) &=& i[H^R,V_k^R(\vec{x},t)]=
R_{kj}\Pi_j^R~,\nonumber\\
\dot{\Pi}_k^R(\vec{x},t) &=& i[H^R,\Pi_k^R(\vec{x},t)]=
(\vec{\partial}^2-M^2)R_{ij}^{-1}V_j^R+J_k~,
\end{eqnarray}
from which we obtain directly the Klein-Gordon equations:
\begin{equation}
\label{nonlKGeq}
(\Box + M^2)V_k^R=R_{kj}J_j~.
\end{equation}
Note that in contrast to the case of the local vector field we do not need
here to make use of the conservation of the fermion four-current.
For massive QED of course current conservation
is a consequence of the equation of motion of the fermions.

For the operator Dirac equation we obtain
\begin{equation}
\dot\Psi^R=i[H^R,\Psi^R]=\gamma_0(-i\gamma_i\partial_i-e\gamma_i V^R_i +m)
\Psi^R +{e\over 2}\left\{\Psi^R,{1\over \vec{\partial}^2-M^2}J_0\right\}~,
\end{equation}
which agrees with the classical form.
The second term on the right is the contribution from the Yukawa potential.
For the Lorentz transformation properties we find
\begin{eqnarray}
\delta_L V_k^R &=& i\epsilon_i[K_i^R,V_k^R]=
\epsilon_i (x_i\partial_t+t\partial_i)V_k^R
-\epsilon_k{1\over \vec{\partial}^2-M^2}J_0
 - \partial_k\Lambda~,\\
\delta_L^0\Psi^R &=& i\epsilon_i[K_i^R,\Psi^R]=
\epsilon_i(x_i\partial_t+t\partial_i)\Psi^R+{1\over 4}
\epsilon_k[\gamma_0,\gamma_k]\Psi^R+ie\left\{\Lambda,\Psi^R\right\}~,
\end{eqnarray}
with
\begin{equation}
\Lambda = \epsilon_i{1\over\vec{\partial}^2-M^2}
\left(\dot{V}_i^R+{1\over \vec{\partial}^2-M^2}\partial_iJ_0\right).
\end{equation}
These quantum Lorentz transformation properties have the same form as the
corresponding classical ones.
In the massless limit it contains a gauge transformation with the gauge
(operator-) function $\Lambda$.
Again the Poincar\'e algebra is satisfied.

\subsection{Vertices and free propagator for the local vector fields $V_i$}

In the following two sections we shall compare the Feynman rules of the
nonlocal reduced fields with those of the local fields. This will
further elucidate the physical significance of the nonlocal reduced
fields and the possible advantage for theories beyond QED where
the fermionic current might not be conserved.

In this section we shall first give a brief review of the Feynman rules
for the local massive vectorfield.
So far the field operators have been given in the Heisenberg representation.
In order to investigate the Feynman rules in the operator approach
it is necessary to pass to the interaction representation
\begin{equation}
(v_i,\pi_i,\psi,\pi)\equiv e^{iH_0t}(V_i,\Pi_i,\Psi,\Pi)
                                  e^{-iH_0t}
\end{equation}
where the Hamiltonian $H$ is split into a free-particle part $H_0$ and the
interaction $V(t)$
\begin{equation}
H=H_0+V(t)~,
\end{equation}
with
\begin{eqnarray}
H_0 &=& \int d^3\vec{x}\left\{{1\over 2}\left(\pi_iR^{-1}_{ij}\pi_j-
                 v_i(\vec{\partial}^2-M^2)R_{ij}v_j\right)-
\pi\gamma_0(\gamma_i\partial_i+im)\psi\right\}~,\\
V(t) &=&\int d^3\vec{x}\left(
v_ij_i + {1\over M^2}j_0^2 - j_0{1\over M^2}\partial_i\pi_i\right)~.
\end{eqnarray}
The fields $v_k$ and $\pi_k$ then satisfy by definition
the free Heisenberg equations
\begin{eqnarray}
\dot{v}_k=i[H_0,v_k]&=&
\pi_k-{1\over M^2}\partial_k(\partial_i v_i)~,\nonumber\\
\dot{\pi}_k=i[H_0,\pi_k]&=&
(\vec{\partial}^2- M^2)v_k-\partial_k(\partial_i v_i)~.
\label{vpi}
\end{eqnarray}
If one defines
\begin{equation}
\label{v_0}
v_0\equiv -{1\over M^2}\partial_i\pi_i~,
\end{equation}
then the two Heisenberg equations (\ref{vpi}) can be combined to
\begin{equation}
(\Box + M^2)v_k+\partial_k \partial_{\nu}v^{\nu} = 0~.
\end{equation}
Taking the divergence of this equation one obtains the covariant
operator equations
\begin{equation}
\partial_{\mu}v^{\mu}=0~,~~~~~~~(\Box -M^2)v^{\mu}=0~.
\end{equation}
These are solved by
\begin{equation}
\label{aa+}
v_{\mu}(\vec{x},t)=\int {d^3\vec{q}\over (2\pi)^3}
{1\over\sqrt{2\omega(\vec{q})}}\sum_{\lambda=1,2,3}
\left(a^{(\lambda)}(\vec{q})\epsilon_{\mu}^{(\lambda)}(\vec{q})
\exp\left[-i(\omega(\vec{q})t- q_i x_i)\right] + h.c.\right)~.
\end{equation}
The annihilation operators $a^{(\lambda)}(\vec{q})$ with polarization
$\lambda = 1,2,3$ satisfy the canonical commutation relations
\begin{equation}
[a^{(\lambda)}(\vec{q}),a^{(\lambda')}(\vec{q}')^+]=\delta_{\lambda\lambda'}
(2\pi)^3\delta(\vec{q}-\vec{q}')
\end{equation}
and the real covariant polarization vectors satisfy the orthonormality
conditions
\begin{equation}
\epsilon^{(\lambda)}(\vec{q})\cdot\epsilon^{(\lambda')}(\vec{q})
=\delta_{\lambda\lambda'}~~~~~~q\cdot\epsilon^{(\lambda)}(\vec{q})=0
\end{equation}
and hence
\begin{equation}
\sum_{\lambda}\epsilon_{\mu}^{(\lambda)}(\vec{q})
\epsilon_{\nu}^{(\lambda)}(\vec{q})
=-\left(g_{\mu\nu}-{q_{\mu}q_{\nu}\over M^2}\right)~.
\end{equation}
Note that the spatial components of this tensor are the Fourier transform
of the inverse reduction operator $R_{ij}^{-1}$.
The representation (\ref{aa+}) leads to the free propagator
\begin{eqnarray}
D^L_{\mu\nu}(x-y)&=&
\langle 0|Tv_{\mu}(\vec{x},x_0)v_{\nu}(\vec{y},y_0)|0\rangle
\nonumber\\
&=&-i\int {d^4q\over (2\pi)^4} {e^{-iq\cdot (x-y)}\over q^2-M^2 +i\epsilon}
\left(g_{\mu\nu}-{q_{\mu}q_{\nu}\over M^2}\right)
-{i\over M^2}\delta_{\mu 0}\delta_{\nu 0}\delta (x-y)~.
\label{D^L}
\end{eqnarray}
The extra noncovariant term is, when sandwiched between two $j_0$, cancelled
by the noncovariant term $j_0^2/M^2$ in V(t). Note that with the definition
(\ref{v_0}) the first and third term in $V(t)$ combine to the covariant
form $j_{\mu}v^{\mu}$.
Alltogether we arrive at the current-current interaction
\begin{equation}
J^{\mu}D_{\mu\nu}(q)J^{\nu}=
-{1\over q^2-M^2}J^{\mu}\left(g_{\mu\nu}-{q_\mu q_\nu \over M^2}
\right)J^{\nu}~.
\label{mPhotprop}
\end{equation}
which is manifestly Lorentz invariant.
However it has a singularity at M=0.
Furthermore the second term leads to ultraviolet
divergences for large momenta. However, since in massive QED the vector field
is coupled to a conserved current $(q_{\mu}J^{\mu}=0)$ the badly
behaved second term drops out and the theory becomes renormalizable.

An alternative, simple way
to obtain the kernel of the current-current interaction is to substitute
the solution of
\begin{equation}
{\delta W_{\rm red}\over \delta V_i}
=-(\partial_t^2-\vec{\partial}^2+M^2)R_{ij}V_j
-{\partial_i\partial_t\over \vec{\partial}^2-M^2}J_0
-J_i = 0~,
\end{equation}
obtained from (\ref{WredmQED}), which has the form
\begin{equation}
V_i=-{R^{-1}_{ij}\over \partial_t^2-\vec{\partial}^2+M^2}
\left({\partial_j\partial_t\over \vec{\partial}^2-M^2}J_0+J_j\right)~,
\end{equation}
into the reduced Lagrangian (\ref{redL}),
leading to the doubly reduced Lagrangian
\begin{equation}
\label{Ldoubll}
{\cal L}_{red^2}= -{1\over 2}J^{\mu}D_{\mu\nu}J^{\nu}
                  +\bar\Psi(i\not{\!\partial}-m)\Psi~,
\end{equation}
with the current-current interaction kernel
\begin{equation}
D^{\mu\nu}=-{1\over \Box+M^2}
\left(g^{\mu\nu}+{\partial^{\mu}\partial^{\nu}\over M^2}\right)~.
\end{equation}
It agrees with (\ref{mPhotprop}).
This substitution corresponds to the
integration over the vector fields in the conventional representation of the
generating functional for connected Greenfunctions.

Finally we note that the corresponding discussion of the fermion field is
standard and can be found e.g. in \cite{Wei}. We shall omit it here because
our main interest is in the vector field rather than the fermion field.

\subsection{Vertices and free propagator for the nonlocal $V^R_k$}

In this section we shall carry out the same steps
for the nonlocal reduced theory.
For the nonlocal fields the interaction-picture Hamiltonian is split into
\begin{equation}
H^R=H_0^R+V^R(t)
\end{equation}
with
\begin{eqnarray}
H_0^R &=& \int d^3\vec{x}\left\{{1\over 2}\left(\pi^R_iR_{ij}\pi^R_j-
                 v^R_i(\vec{\partial}^2-M^2)R^{-1}_{ij}v^R_j\right)+
\pi^R\gamma_0(\gamma_i\partial_i+im)\psi^R\right\}~,\\
V^R(t) &=&\int d^3\vec{x}
\left(v^R_ij_i - {1\over 2}j_0{1\over \vec{\partial}^2-M^2}j_0\right)~.
\end{eqnarray}
The vector field part of the free Hamiltonian $H_0$
 is diagonalized in terms of the nonlocal $V^R$
\begin{equation}
v_{i}^R(\vec{x},t)=\int {d^3\vec{q}\over (2\pi)^3}
{1\over\sqrt{2\omega(\vec{q})}}\sum_{\lambda=1,2,3}
\left(a^{(\lambda)}(\vec{q})\epsilon_{i}^{P(\lambda)}(\vec{q})
\exp\left[-i(\omega(\vec{q})t- q_i x_i)\right] + h.c.\right)~,
\end{equation}
with the real nonlocal polarization vectors defined via
\begin{equation}
\epsilon_i^{P(\lambda)}(\vec{q})\equiv
R_{ij}(q)\epsilon_j^{(\lambda)}(\vec{q})~,
\end{equation}
where $R_{ij}(q)\equiv \delta_{ij}-q_iq_j/(\vec{q}^2+M^2)$ is the Fourier
transform of the reduction operator $R_{ij}$
and hence
\begin{equation}
\sum_{\lambda}\epsilon_{i}^{P(\lambda)}(\vec{q})
\epsilon_{j}^{P(\lambda)}(\vec{q})
= R_{ik}(q)\left(\delta_{kl}+{q_k q_l\over M^2}\right)R_{lj}(k)=R_{ij}(q)=
\delta_{kl}-{q_k q_l\over \vec{q}^2+M^2}~.
\end{equation}
This leads to the free propagator
\begin{eqnarray}
D^R_{ij}(x-y)&=&
\langle 0|Tv^R_{i}(\vec{x},x_0)v^R_{j}(\vec{y},y_0)|0\rangle \nonumber\\
   &=&-i\int {d^4q\over (2\pi)^4} {e^{-iq\cdot (x-y)}\over q^2-M^2 +i\epsilon}
\left(\delta_{kl}-{q_k q_l\over \vec{q}^2+M^2}\right)~.
\end{eqnarray}
Together with the Yukawa potential in the interaction $V^R(t)$ this
leads to the following current-current interaction
\begin{equation}
D^R_{\mu\nu}(q)J^\mu J^\nu = J_0 {1\over \vec{q}^2+M^2} J_0
+J_i J_j\left(\delta_{ij}-{q_i q_j \over \vec{q}^2+M^2}\right)
{1\over q^2-M^2}~.
\label{mvecprop}
\end{equation}
It is the generalization of the photon propagator in Coulomb gauge QED.
It contains the instantaneous Yukawa potential as a generalization of the
Coulomb potential.
In contrast to the conventional covariant massive vector propagator
(\ref{mPhotprop}) the reduced propagator
(\ref{mvecprop}) is regular in the limit $M\rightarrow 0$ and is
well behaved for large momenta.

Before we turn to a further discussion of the propagator for the
nonlocal reduced fields and its relation to the corresponding
standard propagator of the local fields, we quote an alternative derivation
of it. Again, in order to obtain the kernel of the current-current
interaction one can substitute
the solution of
\begin{equation}
{\delta W_{\rm red}\over \delta V^R_i}
=-(\partial_t^2-\vec{\partial}^2+M^2)R^{-1}_{ij}V^R_j-J_i = 0~,
\end{equation}
obtained from (\ref{WredmQED}), which has the form
\begin{equation}
V^R_i=-{R_{ij}\over \partial_t^2-\vec{\partial}^2+M^2} J_j
\end{equation}
into the reduced Lagrangian (\ref{redL}),
leading to the doubly reduced Lagrangian
\begin{equation}
\label{Ldoubl}
{\cal L}_{red^2}= {1\over 2}J_i{R_{ij}\over
                          \partial_t^2-\vec{\partial}^2+M^2}J_j
                   +{1\over 2}J_0{1\over \vec{\partial}^2-M^2}J_0
                  +\bar\Psi^R(i\not{\!\partial}-m)\Psi^R~.
\end{equation}
The effective momentum space propagator for massive vector fields
following from the doubly reduced Lagrangian (\ref{Ldoubl}) is therefore
again (\ref{mvecprop}).

For a better comparison with the conventional covariant propagator
(\ref{mPhotprop}) we rewrite the propagator (\ref{mvecprop}) in the
alternative form
\begin{equation}
D^R_{\mu\nu}(q)J^\mu J^\nu = -{1\over q^2-M^2}\left(J_{\nu}^2
+{(J_iq_i)^2-(J_0q_0)^2\over (\vec{q}^2+M^2)}\right)
\label{mvecprop2}
\end{equation}
Hence we see that for massive QED, where the vector field is coupled to a
conserved current $(q_{\mu}J^{\mu}=0)$, we find that the effective
current-current interactions mediated by
the propagator of the nonlocal reduced fields
(\ref{mvecprop}) and by the conventional covariant propagator
(\ref{mPhotprop}) coincide
\begin{equation}
J^{\mu}D^R_{\mu\nu}J^{\nu}=J^{\mu}D_{\mu\nu}J^{\nu}
\end{equation}

Finally we note that
the propagator of the nonlocal fermion field $\Psi^R$ agrees with the
standard fermion propagator, since $\Psi^R$, defined in (\ref{psired}),
differs from the local $\Psi$ only by a phase shift.

In summary we have found in this section that for massive QED,
where the massive vector field is coupled to a conserved current,
quantization in terms of the nonlocal reduced fields leads to a
quantum theory which is equivalent to the conventional one, but has
improved formal properties. For example the corresponding vector propagator
is well behaved both in the massless and in the ultraviolet limit.

\section{Massive vector theory coupled to a classical current}

We have seen in the preceeding section
that quantization in terms of the nonlocal reduced fields
and conventional quantization in terms of the local vector fields lead
to equivalent quantum theories, when coupled to a conserved current,
as is the case for massive QED.
On the other hand we observe that for the case of nonlocal fields current
conservation is used - via the dynamics of the fermions - only initially
in order to obtain the reduced Lagrangian (\ref{redL}) in terms of the
nonlocal reduced fields from the original Lagrangian (\ref{LQEDloc}).
Namely it was necessary to cancel the
second term in (\ref{Lredpsi}) by a nonlocal phase transformation of the
fermion field. Once the reduced Lagrangian has been obtained, however,
current conservation has not been made use of anymore explicitly,
e.g. for the derivation of the operator Klein-Gordon
equation (\ref{nonlKGeq}), in contrast to the corresponding local case
(\ref{locKGeq}). Furthermore, the nonlocal propagator (\ref{mvecprop}) is
well-defined even if coupled to a nonconserved current in contrast
to the covariant local propagator (\ref{mPhotprop}).

We shall now show that starting from the reduced Lagrangian (\ref{redL})
in terms of the nonlocal fields and coupling it to a nonconserved
classical current instead of the conserved fermion current of QED,
we may obtain a quantum theory which is inequivalent to the corresponding
one in terms of the local field operators.
Let us consider the following Lagrangian of the reduced nonlocal
vector fields $V^R_i$ coupled to some classical current $J_{\mu}^{\rm cl}$
\begin{equation}
{\cal L}_{\rm red}[J_{\mu}^{\rm cl}]
= {1\over 2}\left(\dot{V}^R_iR^{-1}_{ij}\dot{V}^R_j+
          V^R_i(\vec{\partial}^2-M^2)R^{-1}_{ij}V^R_j\right)
      +J_i^{\rm cl}V^R_i+{1\over 2}J_0^{\rm cl}{1\over \vec{\partial}^2-M^2}
                   J_0^{\rm cl}~.
\label{redLJ}
\end{equation}
It is obtained from the reduced Lagrangian (\ref{redL}) of massive QED
in terms of the reduced nonlocal fields by replacing
the coupling to the conserved fermion current of QED by the coresponding
coupling to the classical current. If the classical current is not conserved,
it will lead to a quantum theory of reduced fields, which is inequivalent
to the theory resulting from the Lagrangian
\begin{equation}
\label{LQEDlocJ}
{\cal L}(x)[J_\mu^{\rm cl}]=
-{1\over 4}F_{\mu\nu}F^{\mu\nu}+{1\over 2}M^2V_\mu^2
-V^{\mu}J_{\mu}^{\rm cl}~.
\end{equation}
In particular the results for the vacuum-to-vacuum amplitudes will be
different.
For the number $<0|N|0>$ of massive vector bosons created by the
classical current $J_{\mu}^{\rm cl}$ we obtain
\begin{equation}
\label{Ncreat1}
<0|N|0>_{\rm loc}=\int {d^3\vec{q}\over (2\pi)^3}{1\over 2\omega(\vec{q})}
\left[|J_{\mu}^{\rm cl}(\vec{q})|^2
- {1\over M^2}\left|J_{\mu}^{\rm cl}(\vec{q})q^\mu\right|^2\right]
\end{equation}
for the local theory (\ref{LQEDlocJ}) using (\ref{mPhotprop}), and
\begin{equation}
\label{Ncreat2}
<0|N|0>_{\rm nonl}=\int {d^3\vec{q}\over (2\pi)^3}{1\over 2\omega(\vec{q})}
\left[|J_{\mu}^{\rm cl}(\vec{q})|^2
+{1\over \omega(\vec{q})^2} \left(\sum_{i=1}|J_i^{\rm cl}(\vec{q})|^2 q_i^2
-|J_0^{\rm cl}(\vec{q})|^2 \omega(\vec{q})^2\right)\right]
\end{equation}
for the nonlocal theory (\ref{redLJ}) using (\ref{mvecprop}).
When the classical current is not conserved,
 $\omega(\vec{q})J_0^{\rm cl}-q_iJ_i^{\rm cl}\neq 0$,
the two results (\ref{Ncreat1}) and (\ref{Ncreat2}) are different.

\section{Summary and Conclusions }

Herein we have considered reduced phase space quantization of massive vector
theory
based on the elimination of unphysical degrees of freedom at the classical
level by explicitly solving the constraint equation for $V_0$,
which leads to the reduced action
\begin{eqnarray}
W^{\rm red}&\equiv & W|_{{\delta W \over \delta V_0}=0}~.\nonumber
\end{eqnarray}
We have considered the quantization in terms of reduced nonlocal
fields, which reduce to the transverse photon fields in the massless limit.
We have proven the Poincar\'e invariance of the corresponding quantum theory
by a calculation analogous to that by Zumino for the massless case.
The quantum Lorentz transformation properties of the
nonlocal reduced fields have been derived and shown to coincide
with the corresponding classical ones.
Furthermore in the massless limit they smoothly turn into the
Lorentz transformation properties of the photons.
We obtain a theory that differs from the standard theory in terms of the local
fields by the form of the vector particle propagator (\ref{mvecprop}),
which is the $M\neq 0$ generalization of the photon propagator in Coulomb
gauge QED.
It includes an instantaneous Yukawa potential in generalization of the
Coulomb potential, has no mass singularity, and is well behaved for
large momenta.
The two schemes of quantization, local and nonlocal,
coincide when coupled to a conserved current as in QED.
However, in contrast to the quantization in the local vector fields,
the quantization in the nonlocal reduced fields has a well behaved
propagator in the infrared and the ultraviolet region, even if coupled to
a nonconserved current, such as a classical nonconserved current. In
this case the vacuum-to-vacuum amplitudes are different for the reduced
nonlocal and the local massive vector fields.
Other important examples of theories were the vector field is coupled
to a nonconserved current are massive non-Abelian theories. Although
there seems to be general consensus that they are nonrenormalizable
beyond one loop order \cite{Itz}, it might be interesting to extend our
concept of nonlocal reduced fields to the massive non-Abelian case
as an alternative to recent covariant approaches \cite{Ba1} related to
that of St\"uckelberg .
The corresponding preprojection operator for massive non-Abelian theories
of course will be much more complicated.

Comparing our physical, reduced approach with the covariant one by
St\"uckelberg, we first note that reduced phase space
quantization of massive QED in the reduced fields
in the massless limit smoothly turns into the quantization of QED in
the Coulomb gauge,
whereas the covariant St\"uckelberg approach turns into the Lorentz gauge
quantization of QED.
In the St\"uckelberg approach Poincar\'e invariance is manifest, but
all results have to be proven to be independent of spurious
effects due to the ghosts.
We have shown that it is possible to master the complications
of reduced phase space quantization,
the loss of manifest Poincare invariance and the nonlocality of the fields,
which arise for the sake of quantizing only physical degrees of freedom.
The reduced approach might be an important alternative to the covariant
St\"uckelberg approach to massive non-Abelian gauge theories \cite{Ba1},
where the difficulties of disentangling the spurious from the physical
results may be at least comparable to the difficulties of having to prove the
Poincar\'e invariance of the physical result.

From the properties of the reduced fields found in this paper
we see various interesting applications.
A direct application for example might be
the calculation of relativistic boundstates of fermions coupled
via massive vector fields.
Writing the relativistically invariant Bethe-Salpeter equation for QED
into a form of a single time boundstate equation and comparing different
gauges Love \cite{SLo} showed that the Coulomb gauge with its
instantaneous Coulomb potential gives the most effective description
of relativistic boundstates of two fermions.
Correspondingly we expect the choice of the nonlocal reduced fields with
its instantaneous Yukawa potential to be the most effective way to describe
relativistic fermions bound by massive photons.
In general the description in terms of reduced fields should be
the most effective one in all circumstances where the Coulomb gauge
is the most effective one for QED.

Another important application connected with the main aim of the
present paper, namely to find a Lorentz invariant quantum theory of vector
fields with a well defined massless limit, is the possibility
to study the IR behaviour of the corresponding
massless theory, such as the formation of Lorentz and gauge invariant
condensates \cite{Bla,Pav}.
The infrared catastrophe can be controlled by allowing the gauge field
to have a small finite mass.
For the simple example of massive QED, discussed in this paper,
the nonlocal reduced fields coincide in the zero momentum limit with
the three local vector fields, since $R_{ij}(\vec{q}=0)=\delta_{ij}$
and become transverse photons in the high momentum limit.
For a small finite mass we can therefore regard the nonlocal field
as a smooth interpolation between the three gauge invariant zero momentum
components and the two transverse components for the nonzero momentum
components.
After all calculations have been done with the reduced fields,
the massless limit can then be taken in a smooth way.
Of course in the case of QED the vacuum structure is well known to be trivial.
However, as discussed in \cite{Pav}, already in the scenario of the global
colour model of QCD the transition to nonlocal fields as discussed above
allows one to give an explicit construction of a squeezed
condensate and the corresponding effective quark action with interesting
properties.
The present work could be a first step towards the
investigation of more complicated theories such as QCD
with a nontrivial vacuum structure.

\section{Acknowledgements}

We are grateful to D. Blaschke, S.A. Gogilidze, A.M. Khvedelidze, E.A. Kuraev,
C.D. Roberts and G. R\"opke for fruitful discussions.
H.P.P. acknowledges the support
by the Deutsche Forschungsgemeinschaft under grant No. RO 905/11-1,2 and
V.N.P. the support by the
RFFI, Grant No. 96-01-01223 and the Federal Minister of Research and
Technology (BMFT) within the Heisenberg-Landau Program.


\begin{appendix}

\section{Classical Lorentz transformation properties}

In order to investigate the Lorentz transformation properties of the
classical fields $V_k(x)$, we consider
an infinitesimal boost along $\underline{\epsilon}$
\begin{equation}
\delta x_k =\epsilon_k t \hspace{2cm} \delta t = \epsilon_i x_i~.
\end{equation}
The partial derivatives and vectors then transform as
\begin{eqnarray}
\delta (\partial_k) &=&-\epsilon_k \partial_t~, \hspace{2cm}
\delta (\partial_t) =-\epsilon_i\partial_i~,\\
\delta V_k &=&\epsilon_k V_0~, \hspace{2cm}
\delta V_0 =\epsilon_i V_i~.
\end{eqnarray}
From these we can derive the corresponding infinitesimal changes in the
fields.

The local $V_i$ have the standard transformation properties
\begin{eqnarray}
\delta_LV_k &=& \epsilon_i (x_i\partial_t+t\partial_i)V_k - \epsilon_kV_0
\nonumber\\
&=&\epsilon_i (x_i\partial_t+t\partial_i)V_k + \epsilon_k
{1\over \vec{\partial}^2-M^2}\partial_i\dot{V}_i~,
\end{eqnarray}
realized nonlinearly using (\ref{V_0V_i}) for $V_0$.
The non-local $V_i^R$ on the other hand transform as
\begin{equation}
\delta_LV_k^R[V]
=\epsilon_i (x_i\partial_t+t\partial_i)V_k^R - \partial_k
\left({1\over \vec{\partial}^2 -M^2}
\epsilon_i\dot{V}^R_i\right)~.
\end{equation}
The second term originates from the nonlocality of the physical
$V_k^R[V]$ and corresponds to a gauge transformation.
It is different from the second term in the transformation
formula of the local $V_k$.

\end{appendix}

\end{document}